\documentclass[runningheads]{llncs}
\usepackage[T1]{fontenc}
\usepackage[utf8]{inputenc}
\usepackage{graphicx}
\usepackage{hyperref}
\usepackage{color}

\usepackage{amsmath, amssymb, mathtools}
\usepackage{cleveref}
\crefformat{equation}{\textup{#2(#1)#3}}
\usepackage{algorithm}%
\usepackage{algorithmicx}%
\usepackage{algpseudocode}%
\usepackage{xspace}

\usepackage{pgfplots}
\usepgfplotslibrary{groupplots,dateplot}
\usetikzlibrary{patterns,shapes.arrows}
\pgfplotsset{compat=newest}

\newcommand{\seminorm}[2]{\left\lvert#1\right\rvert_{#2}}

\newcommand{\curl}{\operatorname{curl}}
\renewcommand{\div}{\operatorname{div}}
\def\vec#1{\ensuremath{\mathchoice
		{\mbox{\boldmath$\displaystyle\mathbf{#1}$}}
		{\mbox{\boldmath$\textstyle\mathbf{#1}$}}
		{\mbox{\boldmath$\scriptstyle\mathbf{#1}$}}
		{\mbox{\boldmath$\scriptscriptstyle\mathbf{#1}$}}}}
\newcommand{\grav}{g}

\newcommand{\ICON}{\texttt{ICON-O}\xspace}
\newcommand{\SWEET}{\texttt{SWEET}\xspace}
\newcommand{\OpenMP}{\texttt{OpenMP}\xspace}
\newcommand{\MPI}{\texttt{MPI}\xspace}
\newcommand{\dSDC}{dSDC\xspace}
\newcommand{\coriolis}{F}

\newcommand{\calGh}{\mathcal{G}}
\newcommand{\PG}{\mathcal{P}}

\newcommand{\hPG}{\widehat{\mathcal{P}}}
\newcommand{\MG}{\mathcal{M}}

\newcommand{\gradh}{{\nabla_h}}
\newcommand{\Grad}[1]{\operatorname{N}\left(#1\right)}
\newcommand{\divh}{\operatorname{div_{h}}}
\newcommand{\Div}[2]{\operatorname{D}\left(#1,#2\right)}
\newcommand{\dt}{{\vartriangle}t}

\begin{document}
\title{Parallel performance of shared memory parallel spectral deferred corrections\thanks{	This project has received funding from the European High-Performance Computing Joint Undertaking (JU) under grant agreement No 955701. 
The JU receives support from the European Union’s Horizon 2020 research and innovation programme and Belgium, France, Germany, and Switzerland. 
This project also received funding from the German Federal Ministry of Education and Research (BMBF) grants 16HPC048 and 16ME0679K.
}}
\author{Philip Freese\inst{1}\orcidID{0000-0002-9838-6321} \and
Sebastian Götschel\inst{1}\orcidID{0000-0003-0287-2120} \and
Thibaut Lunet\inst{1}\orcidID{0000-0003-1745-0780} \and
Daniel Ruprecht\inst{1}\orcidID{0000-0003-1904-2473} \and
Martin Schreiber\inst{2}\orcidID{0000-0002-2390-6716}}
\authorrunning{P. Freese et al.}
\institute{Institute of Mathematics, Hamburg University of Technology, Hamburg, Germany\\
\email{\{philip.freese,sebastian.goetschel,thibaut.lunet,ruprecht\}@tuhh.de} \and	
Univ. Grenoble Alpes / Laboratoire Jean Kuntzmann / Inria, Grenoble, France\\
Technical University of Munich, Germany
\email{martin.schreiber@univ-grenoble-alpes.fr}
}
\maketitle              
\begin{abstract}	
	We investigate the parallel performance of Parallel Spectral Deferred corrections, a numerical approach that provides 
	small-scale parallelism for the numerical solution of initial value problems. 
	The scheme is applied to the shallow-water equation and uses an implicit-explicit splitting that,  in order to be efficient, integrates 
	fast modes implicitly and slow modes explicitly.
	We describe parallel \OpenMP-based implementations of parallel Spectral Deferred Corrections for two well established simulation codes:
	the finite volume based operational ocean model \ICON and 
	the spherical harmonics based research code \SWEET.
	We also develop a performance model and benchmark our implementations on a single node of the JUSUF (\SWEET) and JUWELS (\ICON) system at J\"ulich Supercomputing Centre.
	A reduction of time-to-solution across a range of accuracies is demonstrated.
	For \ICON, we show speedup over the currently used Adams--Bashforth-2 integrator with \OpenMP loop parallelization.
	For \SWEET, we show speedup over serial Spectral Deferred Corrections and a second order implicit-explicit integrator.
	\keywords{parallel-in-time methods, spectral deferred corrections, shared-memory parallelization, shallow-water equations, hyperbolic equation}
\end{abstract}
%
%
%
\section{Introduction}
Parallel-in-time (PinT) integration methods can provide additional concurrency when solving initial value problems on massively parallel computers.
Given the ongoing rapid increase in the number of cores in state-of-the-art high-performance computing systems, spatial parallelization alone is unlikely to be sufficient for translating computing power into application performance.
In the terminology introduced by \cite{Gear1988}, there are parallel-across-the-steps methods like Parareal~\cite{LionsEtAl2001}, MGRIT~\cite{FalgoutEtAl2014_MGRIT} or PFASST~\cite{EmmettMinion2012} and parallel-across-the-method algorithms like Runge--Kutta methods with diagonal Butcher table~\cite{Jackson1991}, iterated Runge--Kutta methods~\cite{VanderHouwen1991} or
parallel Spectral Deferred Corrections (SDC)~\cite{Speck2018}.
While parallel across the step methods can scale to a larger number of cores, they are more complicated to implement and require the use of a coarse model to handle information propagation in time, which is costly in terms of developer time and difficult to derive.
By contrast, parallel-across-the-method algorithms are easier to implement and can theoretically provide optimal speedup.
However, they are limited in the number of parallel processes they can use.
Since SDC-based algorithms have already been successfully used in the context of earth system modeling~\cite{HamonEtAl2020,JiaEtAl2013,RuprechtSpeck2016}, exploring the feasibility of using parallel SDC to speed up an operational model is a logical next step.

The contributions of this paper are the following. The first description of the successful integration of parallel SDC into a research code and an operational ocean model including a description of the necessary mathematical modifications in the method. Second, the development of a performance model and the first demonstration that parallel SDC can provide speedup on a single node of a state-of-the-art high-performance computing system and third an illustration that a nested \OpenMP variant with parallel SDC in the outer and fine-grained parallelism in the inner loop provides better performance than loop-parallelization of the mesh operations alone.

\section{Related work}
First attempts to construct parallel-across-the-method Runge--Kutta integrators with diagonal Butcher tables go back to the 1970s~\cite{Butcher1976,Bickart1977} and the work continued to the 1990s~\cite{IserlesNorsett1990,Jackson1991}.
They were ultimately abandoned as the resulting methods lacked accuracy and stability.
More successful and closely related to parallel SDC were iterated Runge--Kutta methods~\cite{VanderHouwen1991}.
Parallel SDC as a method was first described by \cite{Speck2018}, relying on numerical optimization to find good parameters.
Recently, an analytic approach yielded an improved method that could outperform both the previously published iterated Runge--Kutta method and parallel SDC with numerically optimized parameters~\cite{caklovic2024improving}.
However, in these demonstrations, parallel computational cost is only modeled and neither a parallel implementation is discussed nor are runtimes or speedup measured.
The present paper fills this gap.

There are also a few papers that investigate the application of other variants of SDC to atmospheric modeling.
The first is \cite{JiaEtAl2013}, showing that a serial SDC based on implicit Euler provides a high-order integrator for a variety of meteorological test cases.
One of them is the nonlinear evolution of a barotropic instability by \cite{GalewskyScottPolvani2004} which we also use in this paper.
Stability and accuracy of an implicit-explicit split SDC variant is studied for the linear Boussinesq equations by \cite{RuprechtSpeck2016}.
\cite{HamonEtAl2020} investigate performance of a parallel-across-the-steps variant of SDC for the shallow-water equation on a sphere.

A hybrid version between parallel-across-the-steps and parallel-across-the-method SDC are revisionist integral corrections or RIDC~\cite{OngEtAl2016}, which pipeline updates over a few steps.
Also, parallel rational approximation of exponential integrators (REXI) have been applied to the shallow-water case \cite{GuilhermePeixotoSchreiber2024}.
An attempt to speed up the operational ocean model FESOM2 using the parallel-across-the-steps integration method Parareal was very recently published~\cite{PhilippiEtAl2023,PhilippiEtAl2022} but, because of the complexities of building a coarse model, had only limited success.
Moreover, possible improvements using Parareal and MGRIT in atmospheric modeling has recently been studied by \cite{GuilhermePeixotoSchreiber2024}.
Overviews of parallel-in-time literature are given by \cite{Gander2015_Review} and \cite{OngEtAl2020}.


\section{Models}
We briefly sketch the two frameworks for which we benchmark parallel SDC with a focus on numerical time-stepping.
The first model is the ocean part \ICON of the operational earth system model \texttt{ICON}, the second is \SWEET, a research code design for fast exploration of time stepping methods for geophysical fluid dynamics.

\subsection{\ICON}
The Icosahedral Nonhydrostatic Weather and Climate Model (\texttt{ICON}) \cite{Wan2013,Zaengl2015} is used both for research and operational weather forecasts.
Most relevant for this work, \texttt{ICON} supports \OpenMP for shared memory parallelism.
In this work, we focus on pure, nested \OpenMP parallelization on a single node and discuss extensions in the outlook.

\ICON is the ocean-sea ice component of \texttt{ICON}.
It solves the hydrostatic Boussinesq equations of large-scale ocean dynamics with a free surface, also referred to as the primitive equations of large-scale ocean dynamics \cite{Kornetal2022,Korn2017}.
A special case of this complex dynamical system are the shallow-water equations (SWE), which we use as benchmark model.
For the velocity $\vec{v}$, the fluid thickness $\eta$,
the vertical component of the vorticity
$\omega = \vec{k} \cdot \curl
(\vec{v}_{1}, \vec{v}_{2},  0)^{\top} =
\partial_x \vec{v}_{2} - \partial_y \vec{v}_{1}$
and the bottom topography $b$, the SWE in inviscid, vector invariant form~\cite{KornLinardakis2018} read

\begin{align}
	\begin{split}
		\label{eqsys:SWE_Korn}
		\partial_t \vec{v} + \grav \nabla( \eta + b) &= - (\coriolis + \omega) \vec{k} \times \vec{v} - 	\nabla{\seminorm{\vec{v}}{}^2}/{2}, \\
		\partial_t \eta + \div(\eta \vec{v}) &= 0.
	\end{split}
\end{align}
Here $g$ is the earth's gravity acceleration, $\coriolis$ is the Coriolis parameter and $\vec{k}$ the vector perpendicular to the earth's surface.

Since we focus on time integration and parallel performance, we omit the details of the space discretization.
For readers interested in the particular discretization, we refer to
the full space discrete model given by \cite[equation (40)]{KornLinardakis2018}.
We consider the following space discrete counterpart of~\cref{eqsys:SWE_Korn} where the differential operators mimic discrete counterparts of their continuous originals
\begin{align}
	\begin{split}
		\label{eqsys:timeintegration_sdc}
		\partial_t \vec{v} + \overbrace{\grav \gradh\left(\eta + b\right)}^{\Grad{\eta}} &=
		\overbrace{	- \MG^{-1} \hPG^\dagger (\coriolis + \omega) \hPG \vec{v} - \gradh \seminorm{\PG\vec{v}}{}^2 / 2}^{\calGh(\vec{v})}, \\
		\partial_t \eta + \underbrace{\divh\left(\PG^\top \eta \PG \vec{v}\right)}_{\Div{\eta}{\vec{v}}} &= 0.
	\end{split}
\end{align}
The linear operators $(\PG, \hPG, \hPG^\dagger)$, $\MG=\PG^\top\PG$ build an \textit{admissible reconstruction}~\cite[Definition 6]{KornLinardakis2018}, using a lumped mass matrix $\MG^{-1} \approx  \vec{I}$.

The standard time integration scheme in \ICON is the semi-implicit Adams--Bashforth-2 scheme (AB), which is commonly used in ocean dynamics \cite{CampinAdcroftHillMarshall2004,DanilovSidorenkoWangJung2017,KornLinardakis2018}.
Again, the details of the implementation are found in the literature~\cite[Section 5.4.2]{KornLinardakis2018}.
AB first predicts a velocity to be used in the free surface equation that is solved using a conjugate gradient (CG) method.
Eventually, the new velocity is calculated as a correction of the prediction using the gradient of the new fluid thickness.

We use classical notation from time integration and divide the time interval $[0,T]$ in equidistant time slices $[t_{n},t_{n+1}]$ of length $\dt=t_{n+1}-t_{n}$. The approximate solutions at time $t_{n}$ are denoted as $\vec{v}_{n}$ and $\eta_{n}$, respectively, and thus the AB scheme reads:
\begin{subequations}\label{eqsys:AB_scheme}
	\begin{align}
		&\bar{\vec{v}}_{n+1} = \vec{v}_{n} - \dt \tfrac{2}{5}  \Grad{\eta_{n}} + \dt (\tfrac{3}{2} + 0.1) \calGh(\vec{v}_{n}) - \dt (\tfrac{1}{2} + 0.1) \calGh(\vec{v}_{n-1}), \\
		&\eta_{n+1} - (\dt \tfrac{3}{5})^2 \Div{\eta_{n}}{\Grad{\eta_{n+1}}} = \eta_n - \dt \Div{\eta_{n}}{\tfrac{3}{5} \bar{\vec{v}}_{n+1} + \tfrac{2}{5} \vec{v}_n}, \\
		&\vec{v}_{n+1} = \bar{\vec{v}}_{n+1} - \dt \tfrac{3}{5} \Grad{\eta_{n+1}}.
	\end{align}
\end{subequations}
To summarize, the expression $\Grad{\eta}$ is integrated implicitly, whereas the nonlinear part $\calGh(\vec{v})$ uses explicit integration.
For the expression $\Div{\eta}{\vec{v}}$, we use a mixture, i.e., explicit in $\eta$ and implicit in $\vec{v}$.
Let us emphasize that the proposed scheme is a first-order method. The underlying $\Theta$-scheme-parameters are chosen off-centered for stability reasons.
\subsection{\SWEET}\label{sec:modelSWEET}

The \emph{Shallow-Water Equation Environment for Tests, Awesome!}
(\SWEET) \cite{schreiber2018beyond} is
a research-grade software meant for fast prototyping exploration of time integration methods for PDEs using global spectral
methods (Fourier transform and spherical harmonics).
In particular, it can be used to solve the SWE on the rotating sphere:
\begin{equation}
	\label{eq:sweSWEET}
	\left[\begin{array}{c}
		\frac{\partial\Phi}{\partial t}\\
		\frac{\partial{\mathbf{V}}}{\partial t}
	\end{array}\right]=\underbrace{\left[\begin{array}{c}
			-\overline{\Phi}\nabla\cdot{\mathbf{V}}\\
			-\nabla\Phi
		\end{array}\right]}_{L_{g}(U)}+\underbrace{\left[\begin{array}{c}
			0\\
			-\coriolis\vec{{k}}\times{\mathbf{V}}
		\end{array}\right]}_{L_{c}(U)}+\underbrace{\left[\begin{array}{c}
			-{\mathbf{V}}\cdot\nabla\Phi'\\
			-{\mathbf{V}}\cdot\nabla{\mathbf{V}}
		\end{array}\right]}_{N_{a}(U)}+\underbrace{\left[\begin{array}{c}
			-\nabla\Phi'\cdot{\mathbf{V}}\\
			0\\
		\end{array}\right]}_{N_{r}(U)},
\end{equation}
with $\boldsymbol{V}$ the velocity, $\Phi$ the geopotential,
$\vec{k}$ the vector perpendicular to the earth's surface
and $\coriolis$ the Coriolis parameter.
\SWEET considers those equations in their vorticity-divergence formulation
in Fourier space,
which means that the solution vector is
$\widehat{U} \coloneqq (\hat{\Phi'}, \hat{\zeta}, \hat{\rho})^\top$,
with $\Phi'$ the perturbation on the geopotential, $\zeta$ the vorticity and
$\rho$ the divergence.
We refer to \cite{hack1992description} for further details.

The right-hand side of \cref{eq:sweSWEET} is split into different terms or \emph{tendencies}, that each are related to specific physical processes.
$L_{g}(U)$ is the gravitational-related linear part, directly related to a linear wave equation.
It can be treated fully implicit to avoid very restrictive time-step conditions as it models the behavior of fast frequency modes in the solution (gravity waves) where the system of equations is solved directly in spectral space.
$L_{c}(U)$ is the Coriolis effect part and is often integrated explicitly, separately from the gravity modes.
This splitting, also used in the ECMWF model, allows to use an implicit diagonal solver for $L_g(U)$ that is very efficient and easy to parallelize.
The $N_a(U)$ and $N_r(U)$ are non-linear terms that are usually treated explicitly, which requires transforming the solution variables from spectral space, evaluating the tendencies in real space, and transforming back to spectral space again.
The evaluation of those terms is usually the most expensive part in any
spectral-discretization based solver.
In \SWEET, the parallel FFT is done using the \texttt{fftw3} library and the spherical harmonics transformations use the \texttt{SHTns} library \cite{schaeffer_efficient_2013}.
Parallelization of most vector computations, FFTs and implicit linear solves is done with \OpenMP.

Out of the many time-integration algorithms already available in \SWEET,
we choose as a reference method an implicit-explicit (IMEX) time integration scheme of order 2.
It uses a second-order Strang splitting, integrates  $L_{g}(U)$ with a Crank-Nicholson scheme and uses an explicit second-order Runge--Kutta
time integration (Heun's method) to integrate $L_{c}(U) + N_a(U) + N_r(U)$.
We label this time-integration algorithm \texttt{IMEX-o2}.
The motivation is to have a reference time discretization similar to the semi-implicit AB scheme used in \ICON.

\section{Parallel SDC}
Spectral Deferred Corrections with fast-wave slow-wave IMEX splitting~\cite{RuprechtSpeck2016}
integrate an initial value problem of the form
\begin{equation}
	\label{eq:ivp}
	\partial_t u(t) = f_{\text{I}}(u(t)) + f_{\text{E}}(u(t)), \ u(0) = u_0
\end{equation}
where $ f_{\text{I}}$ are the fast, linear dynamics to be integrated implicitly and $f_{\text{E}}$ the slow,
non-linear dynamics treated explicitly.
SDC can be interpreted as an iterative procedure to approximate the solution
of a collocation method, which can be seen as a fully implicit additive Runge--Kutta method
with a dense Butcher table~\cite{HairerEtAl1993_nonstiff}.

\subsection{IMEX SDC}
To summarize the approach, we denote by
$t_{n} \leq t_{n} + \tau_1 < \ldots < t_{n} + \tau_M \leq t_{n+1}$
the quadrature nodes of the collocation method applied on the interval
$[t_n, t_{n+1}]$ ($0 \leq \tau_{m} \leq 1$),
and define $u_{m}^k$ an approximate solution of~\cref{eq:ivp}
at quadrature node $m\in\{1,\dots,M\}$.
Considering a fixed number of iterations $K$,
for $k\in\{0,\dots,K-1\}$ the IMEX SDC iteration (or sweep) reads
\begin{equation}
	\begin{split}
		u_{m}^{k+1} = u_n &+ \dt \sum_{j=1}^M q_{m,j} \left[ f_{I}(u_{j}^k) + f_{E}(u_{j}^k)\right] \\
		&+\dt\sum_{j=1}^{m-1} {\vartriangle}\tau_{j+1} \left[
		f_E(u_{j}^{k+1}) - f_E(u_{j}^{k}) \right] \\
		&+\dt\sum_{j=1}^m {\vartriangle}\tau_j \left[
		f_I(u_{j}^{k+1}) - f_I(u_{j}^{k}) \right],
	\end{split}
	\label{eq:node_update_sdc}
\end{equation}
where $\dt = t_{n+1} - t_n$ is the time step size, $q_{m_j}$ are the entries of
the Butcher table of the collocation method and ${\vartriangle}\tau_{j+1} = \tau_{j+1}-\tau_{j}$, $j=1,\dots,M-1$, the
distance between the nodes (${\vartriangle}\tau_{1}=\tau_{1} - t_{n}$).
A pseudocode of the IMEX SDC is depicted in \cref{alg:IMEX_sdc}, where the initial guess on each node $u_m^0$ is a copy of the solution $u_n$ at the start of the current time step.
Since \cref{eq:node_update_sdc} depends for a given $u_{m}^{k+1}$ on
already computed values for $u_{j}^k,\; j \in \{1,\dots,M\}$
and $u_{j}^{k+1},\; j \in \{1,\dots,m\}$, the internal node update loop
in \cref{alg:IMEX_sdc} is done sequentially over the $M$ quadrature nodes
and cannot be parallelized.
However, some modifications of the algorithm can be done to introduce natural parallelism.
\begin{algorithm}
	\caption{IMEX SDC time-step update $ t_n \rightarrow t_{n+1}$}
	\label{alg:IMEX_sdc}
	\begin{algorithmic}[1]
		\State $ E_{0} \leftarrow f_{E}(u_{n}) $
		\State $ I_{0} \leftarrow f_{I}(u_{n}) $
		\For{$m:1,\dots, M$}
		\Comment{Sweep initialization}
		\State $ u_{m}^{0} \leftarrow u_{0} $; \; $ E_{m}^0 \leftarrow E_{0} $; \; $ I_{m}^0 \leftarrow I_{0} $
		\EndFor
		\For{$ k:0,\dots, K-1 $}
		\Comment{Sweep iterations}
		\For{$ m:1,\dots, M $}
		\Comment{Node updates}
		\State $ u_{m}^{k+1} \leftarrow $ update \cref{eq:node_update_sdc}
		\State $ E_{m}^{k+1} \leftarrow f_{E}( u_{m}^{k+1}) $; \; $ I_{m}^{k+1} \leftarrow f_{I}(u_{m}^{k+1}) $
		\EndFor
		\EndFor
		\State $ u_{n+1} \leftarrow u_{M}^{K}$
	\end{algorithmic}
\end{algorithm}

\subsection{Diagonal IMEX SDC}

First, \cite{Weiser2014} realized that it is possible to replace the ${\vartriangle} \tau_j$ for the implicit part
with algebraic parameters $q^\vartriangle_{m,j}$ satisfying $q^\vartriangle_{m,j} = 0$ if $j < m$
to speed up convergence without changing the sweep-structure of SDC.
\cite{Speck2018} then observed that if the parameters are chosen such
that $q^\vartriangle_{m,j} = 0$ if $m \neq j$, the computation of all $u_m^{k+1}$ can be parallelized over $M$ tasks,
as long as it does not depend on the explicit part.
So using any non-zero coefficient $q^\vartriangle_{m,m}$
for the implicit part and zero coefficients for the explicit part,
the sweep update becomes :
\begin{equation}
	\begin{split}
		u_{m}^{k+1} = u_n &+ \dt \sum_{j=1}^M q_{m,j} \left[ f_{I}(u_{j}^k) + f_{E}(u_{j}^k)\right] \\
		&+\dt q^\vartriangle_{m,m}\left[
		f_I(u_{m}^{k+1}) - f_I(u_{m}^{k}) \right],
	\end{split}
	\label{eq:node_update_psdc}
\end{equation}

In practice, a good choice of the diagonal coefficients is not a trivial task.
In this paper, we use the recently introduced \texttt{MIN-SR-FLEX} diagonal coefficients for the fast linear term (varying between sweeps)~\cite{caklovic2024improving} and zero coefficients for the slow non-linear term.
The aforementioned work gives theoretical results on the optimal choice of parameters and also proposes further coefficients.
For our setup, the \texttt{MIN-SR-FLEX} parameters performed best regarding stability and accuracy,
that is $q^\vartriangle_{m,m} = \tau_m/k$ at sweep $k$.
Furthermore, we use SDC with $M=4$ Radau nodes such that $\tau_M=t_{n+1}$, i.e., the last quadrature node coincides with the next time point, and always perform $K=M$ iterations.
While this choice might not be the optimal one, it still gives a good balance between stability and accuracy as elaborated in \cite{caklovic2024improving}.

Implementation of diagonal IMEX SDC (\dSDC) requires extension of
existing code, which can be straightforward if those routines are available :
\begin{enumerate}
	\item evaluation of the explicit right-hand-side (or tendency) : $f_E$
	\item evaluation of the implicit right-hand-side : $f_I$
	\item a solver for computing the value of $u_m^{k+1}$ for :
	\begin{equation}
		u_m^{k+1} - \alpha f_I(u_m^{k+1}) = \beta
	\end{equation}
	with scalar $\alpha$ and right-hand-side vector $\beta$, used to solve \eqref{eq:node_update_psdc}.
\end{enumerate}
In particular for constant time-steps,
then $\alpha$ for one node $m$ and sweep $k$ stays
constant over each time-step.
Pseudocode for \dSDC is given in \cref{alg:par_sdc},
where we identify the loop on $m$ that can be executed in
parallel,
as update \eqref{eq:node_update_psdc} depends only on $u_m^{k+1}$
and the quadrature node values $(u_j^{k})_{1 \leq j \leq M}$
already computed during sweep $k$.
Note that in the last sweep, we only need to compute the contribution of the
last node, as all other values are not relevant.
Eventually, since $\tau_{M}=t_{n+1}$,
the iterates of the next time step are given
as the final sweep iterates, i.e.,
we simply take $u_M^{K}$ as the final step solution and do not need to
evaluate the tendencies for $u_M^{K}$.
\begin{algorithm}
	\caption{Diagonal IMEX SDC (\dSDC) time-step update $ t_n \rightarrow t_{n+1}$}
	\label{alg:par_sdc}
	\begin{algorithmic}[1]
		\State $ E_{0} \leftarrow f_{E}(u_{n}) $
		\State $ I_{0} \leftarrow f_{I}(u_{n}) $
		\For{$m:1,\dots, M$}
		\Comment{Sweep initialization}
		\State $ u_{m}^{0} \leftarrow u_{0} $; \; $ E_{m}^0 \leftarrow E_{0} $; \; $ I_{m}^0 \leftarrow I_{0} $
		\EndFor
		\For{$ k:0,\dots, K-2 $}
		\Comment{Sweep iterations}
		\State \textit{Parallel for loop}
		\For{$ m:1,\dots, M $}
		\Comment{Node updates}
		\State $ u_{m}^{k+1} \leftarrow $ update \eqref{eq:node_update_psdc}
		\State $ E_{m}^{k+1} \leftarrow f_{E}( u_{m}^{k+1}) $; \; $ I_{m}^{k+1} \leftarrow f_{I}(u_{m}^{k+1}) $
		\EndFor
		\EndFor
		\State $ u_{M}^{K} \leftarrow $ update \eqref{eq:node_update_psdc}
		\Comment{Last node update}
		\State $ u_{n+1} \leftarrow u_{M}^{K}$
	\end{algorithmic}
\end{algorithm}

Evaluation of the right-hand-side $f_I$ and $f_E$ were straightforward within
both \ICON and \SWEET, but retrieving the solver routine in \ICON
requires some work.
Note that \cref{alg:par_sdc} corresponds to a time-sequential algorithm,
as it does not imply any loop parallelization yet.
Different parallel implementation approaches can be considered,
and the one we chose will be discussed and detailed in a following section.

\subsection{Update in \ICON}

Because \ICON does not directly rely on the form~\cref{eq:ivp} of the initial value problem with
$u = (\vec{v}, \eta)^\top$,
some modifications to the SDC formulation \cref{eq:node_update_sdc} are necessary.
Diagonal IMEX SDC in \ICON mirrors the three-step approach of the already implemented semi-implicit AB scheme~\cref{eqsys:AB_scheme}.
An SDC sweep in \ICON reads
\begin{subequations}
	\label{eq:ICON_update}
	\begin{align}
		&\bar{\vec{v}}_{m}^{k+1} = \vec{v}_{n} + \frac{\tau_{m}\dt}{k + 1} \Grad{\eta_{m}^{k}} + \dt \sum_{j=1}^{M} q_{m,j} \left( \calGh(\vec{v}_{j}^{k}) - \Grad{\eta_{j}^{k}} \right),\\
		\begin{split}
			&\eta_{m}^{k+1} - \left(\frac{\tau_{m}\dt}{k + 1}\right)^2 \Div{\eta_{n}}{ \Grad{\eta_{m}^{k+1}}} \\
			&= \eta_{n} - \frac{\tau_{m}\dt}{k + 1} \left(\Div{\eta_{n}}{\bar{\vec{v}}_{m}^{k+1}} - \Div{\eta_{n}}{\vec{v}_{m}^{k}}\right) - \dt \sum_{j=1}^{M} q_{m,j} \Div{\eta_{j}^{k}}{\vec{v}_{j}^{k}},
			\label{eq:ICON_update_implicit}
		\end{split}\\
		&\vec{v}_{m}^{k+1} = \bar{\vec{v}}_{m}^{k+1} - \frac{\tau_{m}\dt}{k + 1} \Grad{\eta_{m}^{k+1}},
	\end{align}
\end{subequations}
for $m=1,\ldots,M$ and $k=0, \ldots, K-1$.
Here, we used the notation introduced in \cref{eqsys:timeintegration_sdc}.

We use inexact iterative solves for the solution of \cref{eq:ICON_update_implicit} in every step to further improve SDC efficiency~\cite{SpeckEtAl2016}.
The idea is that in early sweeps, implicit solves need not yet be very accurate because the SDC solution is not yet accurate.
As SDC converges and accuracy improves, the previous iterates will provide increasingly accurate starting values for the iterative scheme.
To solve the implicit part of the time-stepping, we use a CG scheme where we restrict the number of CG iterations to a maximum of $4$ (for the established AB scheme, we use a maximum of $2$ CG iterations).
This latter choice ensures load balance, as the individual time step sizes vary and, consequently,
the required steps to reach the tolerance also differ.
The first nodes correspond to smaller time steps and thus require fewer CG iterations.
In contrast, the last node corresponds to the largest step size and could require hundreds of iterations.

The implementation of diagonal SDC in \ICON requires some extensions of the
existing code for the following two reasons.
First, the \ICON implementation is made for fixed time-step sizes, which
conflicts the nature of SDC using different steps.
Second, in \ICON the time-integration is implemented only to be sequential.
In particular, there exist update functions that are called once in a step that
bring internal variables to the next step.
Again this does not match the SDC requirements, as here, during a sweep, we go
back and forth in time.
Consequently, there is a memory footprint in SDC as we need to store previously computed values. 
Moreover, we had to write a new time-step routine, that includes specific
calls of tendency evaluations and the CG solver.

\subsection{Update in \SWEET}

We overload the notation $u:=\widehat{U}$, as the time integration scheme uses the state variables transformed in spectral space.
Using the \texttt{MIN-SR-FLEX} preconditioning of \cite{caklovic2024improving} for the implicit tendency $L_g(u)$,
and zero coefficients for the explicit tendencies $L_c(u) + N_a(u) + N_r(u) := T_E(u)$,
translation of \eqref{eq:node_update_psdc} in \SWEET is :
\begin{equation}
	u_m^{k+1} - \frac{\tau_{m}\dt}{k+1} L_g(u_m^{k+1}) = u_0
	+ \dt\sum_{i=1}^{M} q_{m,j} T_E(u_i^{k}), \quad k \in \{0, \dots, K-1\}.
	\label{eq:SWEET_update}
\end{equation}
Because it is a pseudo-spectral code, \SWEET already provides well-defined
routines for the tendencies evaluation and the Backward Euler solve in \eqref{eq:SWEET_update}.
Hence the implementation of diagonal SDC is quite straightforward in the code.

\subsection{Time-parallel implementation}

Since the parallel part in \cref{alg:par_sdc} corresponds to computing
a for loop in parallel, it is natural to use \OpenMP, although alternative
approaches using \MPI-based time-parallelization, requiring additional
communications between quadrature nodes, are currently investigated.

Since \ICON already uses a hybrid \MPI and \OpenMP parallelization, the additional \OpenMP parallelization in time requires using nested \OpenMP.
Even if \ICON uses a pure \MPI parallelization, the hybrid \MPI and \OpenMP parallel implementation is not straight forward.
The reason is that the global \MPI communication during the CG solves is done
within the outer \OpenMP~
time loop.
This however requires duplicates of the \MPI communicators.
By now, we did not manage to create these communicators within the \ICON framework, which is the main reason we only present a pure \OpenMP parallel version.

For \SWEET, space parallelization of the spherical harmonic transform is
done exclusively with \OpenMP, hence time-parallelization of the diagonal IMEX SDC implementation in the code simply requires using nested \OpenMP.

In the following, we will distinguish the sequential \dSDC version from its time parallel counterpart by noting it either as time-parallel \dSDC or \dSDC(PinT).

\subsection{Space-time parallelization}

Our approach targets overcoming strong-scaling limitations.
Consequently, we choose test cases of small workload where \MPI parallelization would be already close to its limit.
Therefore, the implementation of time-parallel \dSDC in both codes is done using \OpenMP,
as this approach is a natural fitting scheme for time-parallel \dSDC methods
(simple loop parallelization, no need for explicit communication).
For \ICON, each compute node gets assigned a limited number of
additional threads that may be invested in the time parallelization to improve the throughput of the simulation.
For \SWEET,
we expect that the space-time parallelization using two nested \OpenMP
levels uses the compute core resources more efficiently than
the exclusive \OpenMP-based space parallelization originally used
in the code.

For a given number of parallel processes,
one must decide on how \OpenMP binds each thread
on the physical cores of a compute node in the cluster,
especially for nested space-time parallelizations.
Our tests suggest that the best \OpenMP configuration for pure space parallelization is \texttt{OMP\_PLACES=cores} with \texttt{OMP\_PROC\_BIND=close}.
For the time-parallel implementation of \dSDC in \SWEET, the best performing setup was determined as \texttt{OMP\_PLACES=cores} with \newline\texttt{OMP\_PROC\_BIND=close,spread}.
In \ICON, there is not one obvious configuration for all thread numbers.
For less than or equal $48$ threads, the same configuration as for \SWEET gives good results.
However, for more than $48$ threads the runtime deteriorates and the best choice is no binding at all, i.e., \texttt{OMP\_PROC\_BIND=false}.
As the results for smaller thread numbers only differ slightly, for \ICON we stick to the configuration without any binding.

\subsection{Performance model}

We develop an elementary performance model to predict the speedup of \dSDC\xspace compared to time parallel \dSDC, for both \ICON and \texttt{SWEET}.
The computational costs are determined by two main operations:
the explicit evaluation of the tendencies with cost $c_{E}$,
and the implicit solve to compute the updated solution on one quadrature node with cost $c_{S}$.
Both operations are present in \ICON and \SWEET, and are performed in every sweep.
For the \ICON model, additional tendency evaluations of divergence terms are required
due to the three-step approach, which generate an additional cost $\tilde{c}_{E}$ during each sweep update.

Therefore, we develop our performance model as follows.
To compute the SDC initial guess ($k=0$, using a copy of the initial time-step solution),
only the tendencies evaluation with cost $c_{E}$ is required (\cref{alg:par_sdc}, Lines 1-5).
Then, within the sweeps ($k=1,\dots,K-2$), we need one explicit tendency evaluation $c_{E}$
and one implicit solve $c_{S}$ per quadrature node, which can be parallelized (\cref{alg:par_sdc}, Lines 6-12).
For the \ICON model, due to the three-step approach,
each quadrature node requires an additional evaluation of (explicit) divergences,
and we denote this supplementary cost as $\tilde{c}_{E}$
(while for \SWEET we have $ \tilde{c}_{E} = 0 $).
For the last sweep ($k=K$), we only perform the implicit solve (cost $c_{S}$)
for the last quadrature node without any tendency evaluation (\cref{alg:par_sdc}, Line 13).
\Cref{tab:costs} summarizes the costs for \dSDC and time-parallel \dSDC.
\begin{table}
	\centering
	\begin{tabular}{c|c|c}
		stage & \dSDC & \dSDC (PinT)\\
		\hline
		$ k = 0 $ & $ c_{E} $ & $ c_{E} $ \\
		$ k = 1,\dots,K-1 $ & $ M (c_{E} + c_{S} + \tilde{c}_{E}) $ & $ c_{E} + c_{S} + \tilde{c}_{E} $ \\
		$ k = K $ & $ c_{S} $ & $ c_{S} $
	\end{tabular}
	\caption{Computational costs for \dSDC and \dSDC (PinT)}.
	\label{tab:costs}
\end{table}
Hence, the theoretical speedup for parallel over serial \dSDC using $K=M$ sweeps is
\begin{align*}
	S = \frac{C_\text{seq}}{C_\text{par}} = \frac{1 + M (M-1) (1 + \tfrac{\tilde{c}_{E}}{c_{E} + c_{S}})}{1 + (M-1) (1 + \tfrac{\tilde{c}_{E}}{c_{E} + c_{S}})}.
\end{align*}
For \ICON, measurements suggest a value $ \tfrac{\tilde{c}_{E}}{c_{E} + c_{S}} \approx 0.3$,
which yields $S\approx3.39$ with $M=4$.
For \SWEET, since $\tilde{c}_{E}=0$, $S=3.25$, which does not depend on $c_{E}$ and $c_{S}$.
Note that since $\lim_{\tilde{c}_{E}\to\infty} S = M$, the maximum optimal speedup of \dSDC with this configuration
is equal to the number of quadrature nodes $M$.

This is a very elementary performance model that does not take into account memory and communication costs.
But in practice, \dSDC requires storage of  $2 \cdot M$ state vectors in memory, which does not change with the time parallel version.
Hence, the additional cost overhead for memory and time communication simply corresponds to the \OpenMP overhead.
In comparison, the AB scheme in \ICON or the \texttt{IMEX-o2} in \SWEET
require the storage of $2$ state vectors,
so using \dSDC with $M=4$ quadrature nodes increases the memory requirement
by a factor of 4.


\section{Results}
\subsection{Numerical Benchmark}
\begin{figure}[t]
	\centering
	\includegraphics[width=\textwidth]{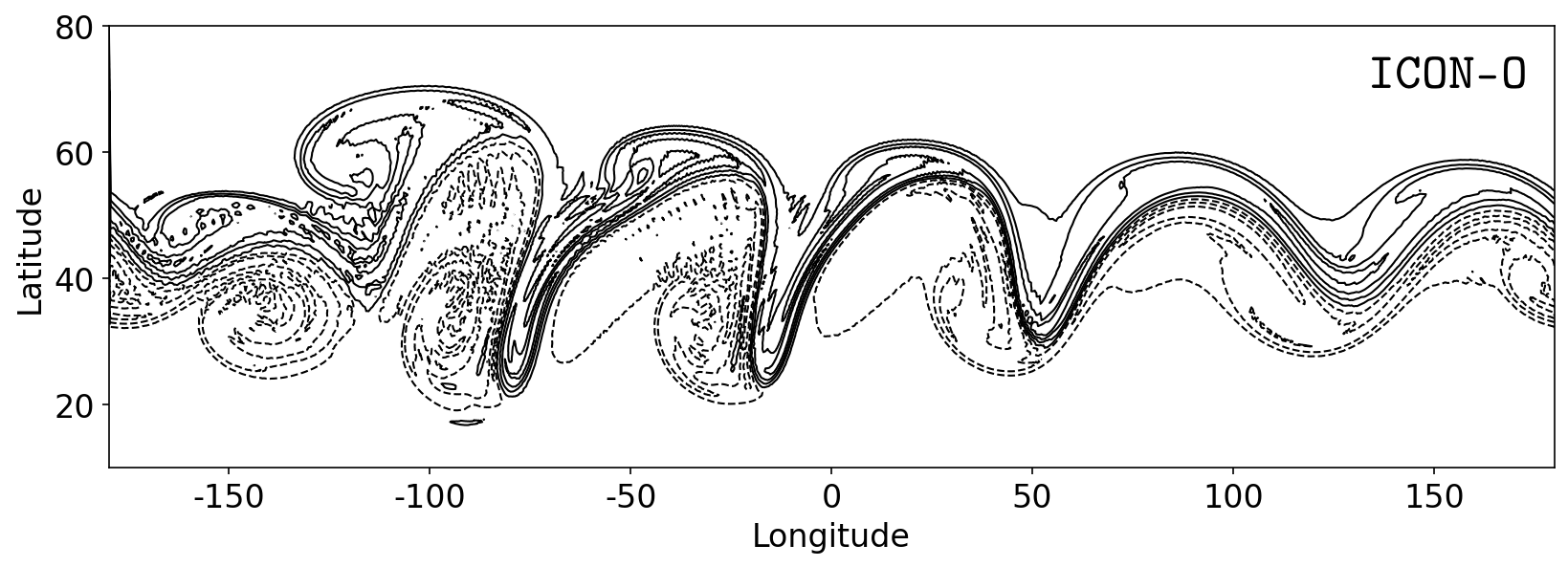}\newline
	\includegraphics[width=\textwidth]{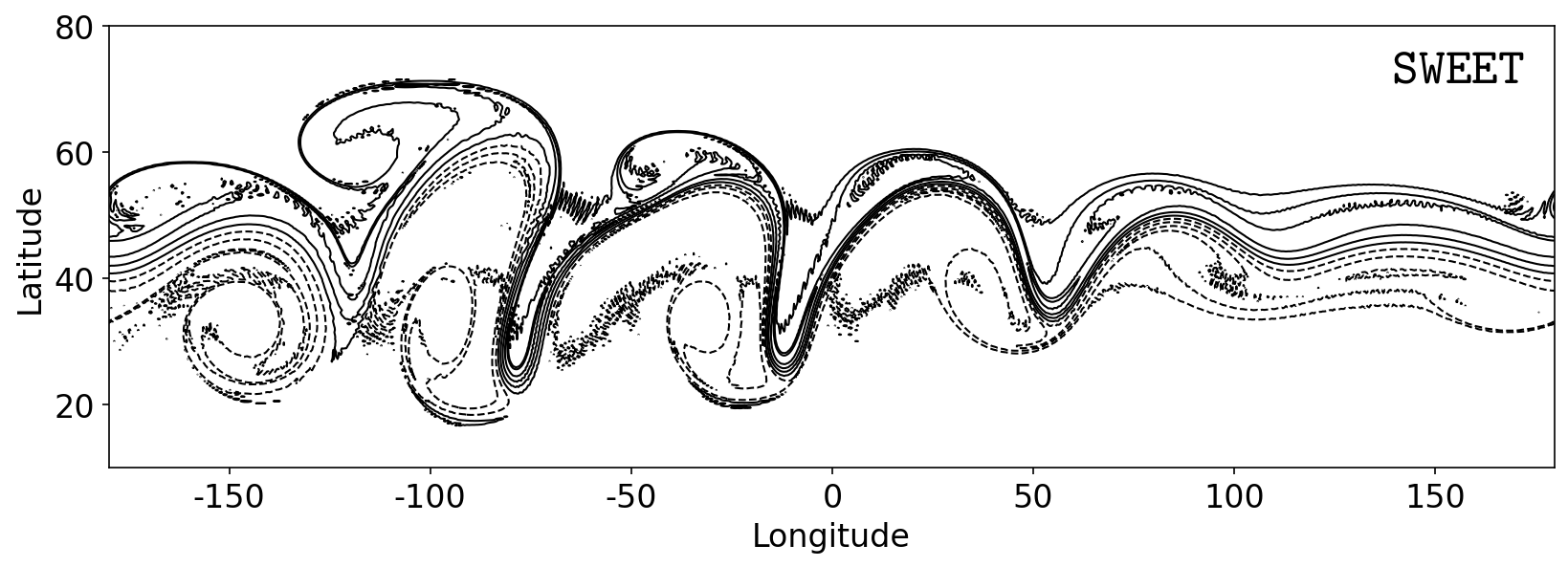}
	\caption{Vorticity contours for the Galewesky test case at the end of day 6.}
	\label{fig:ICON_Galewsky_vort_day6}
\end{figure}

For the numerical experiments and performance demonstration, we choose the well-established Galewsky test case.
The test starts with a geostrophically balanced solution and adds a perturbation that eventually develops into characteristic turbulence.
For the detailed formulation, including the analytical expression of the initial state, we refer to the original work by \cite{GalewskyScottPolvani2004}.
Experiments are run until the end of day 6, until the onset of the transition to fully turbulent flow.
A picture of the vorticity computed with \ICON and \SWEET is shown in~\cref{fig:ICON_Galewsky_vort_day6}.
The fields generally look very similar but with some noticeable differences near the right boundary, due to the different discretizations used in \ICON and \SWEET.
Although the problem size was chosen such that the number of degrees of freedom for both models is similar,  this does not produce the same solution quality, as the spectral space discretization in \SWEET provides a higher spatial accuracy for the same grid resolution.

To evaluate how the time integration methods compare between the two frameworks, we make the hypothesis that the cost of computing one time step is constant whatever its size.
This makes sense considering that we know in advance for both models the required operations to compute the next time step solution (cf. model description).
Before comparing \dSDC and time-parallel \dSDC with the reference time integration methods,
we also determine how to optimally distribute computing cores between
space and time parallel processes, such that we obtain the lowest
computation wall time.

\subsection{Resutls in \ICON}
The results for \ICON were obtained on the JUWELS cluster with the following specifications: 2 x Intel Xeon Platinum 8168 CPU, 2 x 24 cores, 2.7 GHz (96/48 logical/physical cores per compute node).
\ICON was compiled using the new Intel(R) \texttt{oneAPI} C/C++ and Fortran compiler, version 2023.2.0.
\subsubsection{Strong scaling of parallel SDC}
We first investigate strong scalability of time-parallel \dSDC by running simulations on one compute node
of JUWELS for \ICON,
increasing the number of \OpenMP threads used in space and time while keeping
the problem size fixed.
\begin{figure}[t]
	\centering
\begin{tikzpicture}

\definecolor{color0}{rgb}{0.12156862745098,0.466666666666667,0.705882352941177}
\definecolor{color1}{rgb}{1,0.498039215686275,0.0549019607843137}
\definecolor{color2}{rgb}{0.172549019607843,0.627450980392157,0.172549019607843}

\begin{axis}[
legend cell align={left},
width=.49\textwidth,
legend style={
  fill opacity=0.8,
  draw opacity=1,
  text opacity=1,
  at={(.01,0.01)},
  anchor=south west,
  draw=white!80!black
},
log basis x={10},
log basis y={10},
tick align=outside,
tick pos=left,
x grid style={white!69.0196078431373!black},
xlabel={Number of \OpenMP threads},
xmin=0.7, xmax=120,
xmode=log,
xtick style={color=black},
xtick={1, 4, 16, 48, 96},
xticklabels={
    $1$, $4$, $16$, $48$, $96$
},
xmajorgrids,
y grid style={white!69.0196078431373!black},
ylabel={$t_{\textbf{wall}}/(N_{\textbf{steps}} N_{\textbf{dofs}})$  (s)},
ymin=8e-08, ymax=5e-05,
ymode=log,
ytick style={color=black},
ytick={1e-08,1e-07,1e-06,1e-05,0.0001,0.001},
yticklabels={
    \(\displaystyle {10^{-8}}\),
    \(\displaystyle {10^{-7}}\),
    \(\displaystyle {10^{-6}}\),
    \(\displaystyle {10^{-5}}\),
    \(\displaystyle {10^{-4}}\),
    \(\displaystyle {10^{-3}}\)
},
ymajorgrids,
]

\addplot [color0, mark=x, mark size=3, line width=1]
table {%
	1 2.44173418457122e-06
	2 1.35311212414105e-06
	4 8.10921003048556e-07
	8 5.27849489678855e-07
	12 4.28370541294978e-07
	16 3.82643917163645e-07
	24 3.43380187275049e-07
	32 3.44738160539264e-07
	48 3.43491105308321e-07
	64 4.06571984775757e-07
	96 4.27915947958218e-07
};
\addlegendentry{AB}

\addplot [color1,  mark=*, mark size=2.5, mark options={solid}, line width=1]
table {%
	4 9.62487324057007e-06
	8 6.16135452538512e-06
	12 4.96878835843204e-06
	16 4.50481482385695e-06
	24 3.71381699261303e-06
	32 3.47559346922758e-06
	48 3.43148344814584e-06
	64 3.52975077163319e-06
	96 3.98113720219163e-06
};
\addlegendentry{\dSDC (PinT)}

\addplot [color2, dashed, mark=square*, mark size=2, mark options={solid}]
table {%
	1 3.24308689220464e-05
	2 1.86828468890761e-05
	4 1.17770266024133e-05
	8 8.10103082599451e-06
	12 6.81627376329589e-06
	16 6.22013789707971e-06
	24 5.76289736227359e-06
	32 5.91759518090373e-06
	48 6.11282549990574e-06
	64 6.90231939804699e-06
	96 7.17524674856128e-06
};
\addlegendentry{\dSDC}


\end{axis}

\end{tikzpicture}%
\begin{tikzpicture}

\definecolor{color0}{rgb}{0.12156862745098,0.466666666666667,0.705882352941177}
\definecolor{color1}{rgb}{1,0.498039215686275,0.0549019607843137}
\definecolor{color2}{rgb}{0.172549019607843,0.627450980392157,0.172549019607843}

\begin{axis}[
legend cell align={left},
width=.49\textwidth,
legend style={
	fill opacity=0.8,
	draw opacity=1,
	text opacity=1,
	at={(-0.3,1.05)},
	anchor=south west,
	draw=white!80!black
},
log basis x={10},
log basis y={10},
tick align=outside,
tick pos=left,
x grid style={white!69.0196078431373!black},
xlabel={Number of \OpenMP threads},
xmin=0.7, xmax=120,
xmode=log,
xtick style={color=black},
xtick={1, 4, 16, 48, 96},
xticklabels={
	$1$, $4$, $16$, $48$, $96$
},
xmajorgrids,
y grid style={white!69.0196078431373!black},
ymin=8e-08, ymax=5e-05,
ymode=log,
ytick style={color=black},
ytick={1e-08,1e-07,1e-06,1e-05,0.0001,0.001},
yticklabels={
	\(\displaystyle {10^{-8}}\),
	\(\displaystyle {10^{-7}}\),
	\(\displaystyle {10^{-6}}\),
	\(\displaystyle {10^{-5}}\),
	\(\displaystyle {10^{-4}}\),
	\(\displaystyle {10^{-3}}\)
},
ymajorgrids,
]

\addplot [color0, mark=x, mark size=3, mark options={solid,rotate=270}, line width=1]
table {%
	1 2.35855534635082e-06
	2 1.30894476271594e-06
	4 7.86503415202347e-07
	8 5.13231804047835e-07
	12 4.19624763443806e-07
	16 3.77176729516245e-07
	24 3.4166960807849e-07
	32 3.61287482797968e-07
	48 3.83205352250595e-07
	64 4.57642539517204e-07
	96 5.07751792763663e-07
};

\addplot [color1,  mark=*, mark size=2.5, mark options={solid}, line width=1]
table {%
	4 8.97205359257993e-06
	8 5.82709674910983e-06
	12 4.64831496942966e-06
	16 4.14046442321429e-06
	24 3.56616113941196e-06
	32 3.35743154902376e-06
	48 3.19694314694559e-06
	64 3.64879825384684e-06
	96 4.39258156159186e-06
};

\addplot [color2, dashed, mark=square*, mark size=2, mark options={solid}]
table {%
	1 3.03719960867073e-05
	2 1.74363502216146e-05
	4 1.08069886814699e-05
	8 7.3747741809482e-06
	12 6.24244512077138e-06
	16 5.73710457931199e-06
	24 5.47987249248174e-06
	32 5.96763441093552e-06
	48 6.86055232763157e-06
	64 7.54322750806077e-06
	96 8.25024593706543e-06
};

\end{axis}

\end{tikzpicture}%
	\caption{Strong scaling tests for \ICON.\\
		Left: using a larger problem size $N_\text{dofs}=163842$.\\
		Right: using a smaller problem size $N_\text{dofs}=40962$.\\
		This figure shows \textit{only the performance of a single time step} with the possibility of SDC method to take larger time steps not taken into account.}
	\label{fig:strong_scaling_ICON}
\end{figure}
We use the configuration with a time step of $960$s for
the \dSDC versions and a time step of $30$s for the AB scheme.
In \cref{fig:strong_scaling_ICON}, we show the runtime for \dSDC scaled by the number of time steps multiplied by the number of degrees-of-freedom leading to the metric ``time-per-DoF update'' for one time step.
Green squares are for \dSDC whereas orange circles are time-parallel \dSDC using 4 \OpenMP threads times an increasing number of threads to parallelize in space.
The $x$-coordinate corresponds to the total number of threads used,
that is $N_{threads,space}$ for space-parallelization only,
or $M \times N_{threads,space}$ for space-time parallelization with $M=4$.

Scaled runtime is observed for two spatial grid sizes, one with a reference resolution for the problem of interest (\cref{fig:strong_scaling_ICON} left), and one with a coarser resolution (\cref{fig:strong_scaling_ICON} right) for comparison.
We observe that space-parallelization only for \dSDC saturates for
both resolution around $N_{threads,space} \simeq 24$ cores,
which is consistent with the parallel saturation observed for
AB.
However, combining with time-parallelization in \dSDC distributes computation
among $M \times N_{threads,space}$ cores, which saturates around $48$ cores,
hence using more compute cores with better efficiency than space-parallelization
only.
Furthermore, using \dSDC (PinT) allows for a greater reduction of wall-clock time compared to \dSDC with purely space parallelization for both resolutions.
This larger efficiency of space-time parallelization compared to purely space-parallelization on the same number of cores is due to \dSDC belonging to the class of direct time-parallelization algorithms which have higher efficiency than usual iterative time-parallel methods
(see \cite{gander2024time} for the classification),
and that the numerical methods used in \ICON rapidly constrain the space-parallel scalability with large number of threads.
Note that for one time-step only,
the time-parallel speedup of \dSDC (PinT) without space parallelization
(4 \OpenMP threads with $N_{threads,space} = 1$),
is very close to the value of $S_{theory} = 3.39$ given by our performance model
($S\simeq3.37$ for the larger problem, $S\simeq3.39$ for the smaller one).
Finally, using more \OpenMP threads than physical cores available particularly
degrades the performance of each scheme in any case.

We also compare with the scaled wall-clock time of the reference AB time
integration schemes \cref{eqsys:AB_scheme}.
For one time-step only, AB allows minimum wall-clock time around 10 times
lower than time-parallel \dSDC,
which it due to the additional computation work required by \dSDC at each sweep.
However, \dSDC using \texttt{MIN-SR-FLEX} coefficients allows to take a way larger time-step than AB while keeping a similar accuracy,
hence reducing the computation time for a given time-window.
This is investigated in the next section.

\subsubsection{Work precision}
\Cref{fig:ICON_work_precision_speedup} (left) shows work (measured in runtime) against precision (measured in realtive $L^\infty$-$L^2$ error of the vorticity) for $48$ \OpenMP threads.
The reference for \ICON is computed with the AB scheme and a time step of $1$s.
For accuracies below 4 percent, we achieve faster time-to-solution with \dSDC compared to AB.
We emphasize that the maximal time-step for the respective methods are the coarsest possible ones, i.e., for larger time-steps the methods get unstable.
As \cref{fig:ICON_work_precision_speedup} (left) indicates, there is a small range of accuracies where the AB method performs best.
Further optimization of the time parallelization possibly makes \dSDC outperform AB across the full range of accuracies.
Note that \dSDC could even be faster than the coarsest possible AB scheme if the user would accept a larger error.
However, this might still be of interest, as meteorologists are usually interested in stability rather than accuracy.

\begin{figure}[t]
	\begin{tikzpicture}

\definecolor{color0}{rgb}{0.12156862745098,0.466666666666667,0.705882352941177}
\definecolor{color1}{rgb}{1,0.498039215686275,0.0549019607843137}
\definecolor{color2}{rgb}{0.172549019607843,0.627450980392157,0.172549019607843}

\begin{axis}[
	legend cell align={left},
	width=.49\textwidth,
	legend style={
		fill opacity=0.8,
		draw opacity=1,
		text opacity=1,
		at={(-0.3,1.05)},
		anchor=south west,
		draw=white!80!black
	},
	log basis x={10},
	log basis y={10},
	tick align=outside,
	tick pos=left,
	x grid style={white!69.0196078431373!black},
	xlabel={Wall-clock time (s)},
	xmajorgrids,
	xmin=172.685219744705, xmax=2346.4261954148,
	xmode=log,
	xtick style={color=black},
	xtick={204, 303, 487, 972, 1975},
	xticklabels={$ 204 $, $ 303 $, $ 487 $, $ 972 $, $ 1975 $},
	y grid style={white!69.0196078431373!black},
	ylabel={relative $L^\infty(0,T,L^2(\Omega))$},
	ymin=0.000319476122285089, ymax=0.455907576332319,
	ymode=log,
	ytick style={color=black},
	ytick={0.0004,0.002,0.012,0.04,0.1,0.25},
	yticklabels={
		$0.0004$,
		$0.002$,
		$0.012$ ,
		$0.04$,
		$0.1$,
		$0.25$
	},
	ymajorgrids
	]

	\addplot [color0, mark=x, mark size=3, mark options={solid,rotate=270}, line width=1]
	table {%
		1975.45219 0.00633564544841647
		972.4885 0.0128985308110714
		489.27498 0.0255470704287291
		248.59414 0.0495351068675518
	};

	\addplot [color1,  mark=*, mark size=2.5, mark options={solid}, line width=1]
	table {%
		965.88336 0.00045787324779667
		487.62772 0.00179449329152703
		303.5994 0.0122266178950667
		245.76769 0.100027047097683
		204.94408 0.250490248203278
	};

	\addplot [color2, dashed, mark=square*, mark size=2, mark options={solid}]
	table {%
		1715.88569 0.00045787324779667
		860.25892 0.00179449329152703
		540.83028 0.0127972280606627
		434.22376 0.100027047097683
		368.14738 0.250490248203278
	};
\end{axis}

\end{tikzpicture}%
\begin{tikzpicture}
	
\definecolor{color0}{rgb}{0.12156862745098,0.466666666666667,0.705882352941177}
\definecolor{color1}{rgb}{1,0.498039215686275,0.0549019607843137}
\definecolor{color2}{rgb}{0.172549019607843,0.627450980392157,0.172549019607843}

\begin{axis}[
legend cell align={left},
width=.49\textwidth,
legend style={
	fill opacity=0.8,
	draw opacity=1,
	text opacity=1,
at={(0.99,0.01)},
anchor=south east,
	draw=white!80!black
},
log basis x={10},
log basis y={10},
tick align=outside,
tick pos=left,
x grid style={white!69.0196078431373!black},
xlabel={Number of \OpenMP threads},
xmin=1.6480407036756, xmax=116.501976906144,
xmode=log,
xtick style={color=black},
xtick={2,4,16,48,96},
xticklabels={$2$,$4$,$16$,$48$,$96$},
xmajorgrids,
y grid style={white!69.0196078431373!black},
ylabel={Speedup},
ymin=0.888735943775632, ymax=14.9061775584752,
ymode=log,
ytick style={color=black},
ytick={1,2,4,6,12},
yticklabels={1,2,4,6,12},
ymajorgrids
]
\addplot [draw=color0, mark=x, mark size=3, mark options={fill=color0}, line width=1]
table{%
	x  y
	2 1
	4 1.66861152648926
	8 2.56344318389893
	12 3.15874242782593
	16 3.536217212677
	24 3.94056630134583
	32 3.92504286766052
	48 3.939293384552
	64 3.32809948921204
	96 3.1620979309082
};
\addlegendentry{AB}
\addplot [draw=color1, mark=*, mark size=2.5, mark options={fill=color1}, line width=1]
table{%
	x  y
	2 2.47001981735229
	4 4.49871826171875
	8 7.02760744094849
	12 8.71431446075439
	16 9.61184597015381
	24 11.6590528488159
	32 12.4581861495972
	48 12.6183290481567
	64 12.2670392990112
	96 10.8761854171753
};
\addlegendentry{\dSDC (PinT)}
\addplot [color2, dashed, mark=square*, mark size=2, mark options={solid}]
table{%
	x  y
	2 2.3176121711731
	4 3.6766140460968
	8 5.34494733810425
	12 6.35238456726074
	16 6.96119499206543
	24 7.51350927352905
	32 7.3170919418335
	48 7.0834002494812
	64 6.273193359375
	96 6.03457832336426
};
\addlegendentry{\dSDC}
\end{axis}
	
\end{tikzpicture}
	\caption{Left: Work-precision plot for \ICON using $48$ threads, i.e., for AB and \dSDC $48$ threads in space and for \dSDC (PinT) $4$ threads in time and $12$ threads in space.
		Various time-step sizes are used, for AB $\dt \in \{120, 60, 30, 15\}$ (s)
		and for \dSDC $\dt \in \{1440, 1200, 960, 600, 300\}$ (s). AB uses a maximum of $2$ CG iterations,
		\dSDC a maximum of $4$.\\
		Right: Speedup of \dSDC in \ICON, comparing AB ($\dt=30$s) and \dSDC ($\dt=960$s) at similar relative accuracy of $\approx 0.012$,
		using AB with 2 \OpenMP threads (base configuration in \ICON) as base method for speedup.
	}
	\label{fig:ICON_work_precision_speedup}
\end{figure}
\subsubsection{Speedup}
\Cref{fig:ICON_work_precision_speedup} (right) shows speedup for the different methods calibrated to provide an accuracy of around $0.012$ in \ICON.
For the AB method, this requires a time-step of $30$ seconds while for \dSDC we can use a time step size of $960$ seconds (factor $32$ larger).
Since AB with two threads for loop parallelization is the default configuration in \ICON, we use that as the reference to compute speedup.
Replacing AB with the \dSDC (space-parallel only)
results in an acceleration that is approximately twice as high.
This is due to the fact that the significantly larger time step that SDC is capable of performing more than compensates for the increased computing costs associated with each step.
Adding more \OpenMP threads to the loop parallelization provides additional speedup, up to around seven for $24$ threads.
However, using nested \OpenMP with the time-parallel \dSDC as outer loop and \ICON's existing loop parallelization as inner provides the best performance.
Speedup for 48 \OpenMP threads increases to more than 12,
which is a factor 3 compared to the maximum speedup that can be obtained with \OpenMP space-parallelization of \ICON with the AB scheme.

However, those result are specific to \ICON, for which space-parallelization
with \OpenMP rapidly saturates due to the numerical methods used in this solver.
To complete this performance assessment of time-parallel \dSDC,
we also investigate its benefit for \SWEET that uses numerical schemes
allowing a highly optimized \OpenMP-base space-parallelization.

\subsection{Results in \SWEET}
For \SWEET, we used the JUSUF system that has the following specifications: 2 x AMD EPIC 7742, 2 x 64 cores, 2.25 GHz (256/128 logical/physical cores per compute node). Again, the Intel(R) compiler was used to compile \SWEET.

\subsubsection{Strong scaling of parallel SDC}
We first investigate strong scalability of time-parallel \dSDC by running simulations on one compute node of JUSUF,
increasing the number of \OpenMP threads used in space and time while keeping
the problem size fixed.
\begin{figure}[t]
	\centering
\begin{tikzpicture}

\definecolor{color0}{rgb}{0.12156862745098,0.466666666666667,0.705882352941177}
\definecolor{color1}{rgb}{1,0.498039215686275,0.0549019607843137}
\definecolor{color2}{rgb}{0.172549019607843,0.627450980392157,0.172549019607843}
\definecolor{color3}{rgb}{0.83921568627451,0.152941176470588,0.156862745098039}
\definecolor{color4}{rgb}{0.580392156862745,0.403921568627451,0.741176470588235}

\begin{axis}[
legend cell align={left},
width=.49\textwidth,
legend style={
  fill opacity=0.8,
  draw opacity=1,
  text opacity=1,
  at={(.01,.01)},
  anchor=south west,
  draw=white!80!black
},
log basis x={10},
log basis y={10},
tick align=outside,
tick pos=left,
x grid style={white!69.0196078431373!black},
xlabel={Number of \OpenMP threads},
xmin=0.7, xmax=320,
xmode=log,
xtick style={color=black},
xtick={1, 4, 16, 64, 256},
xticklabels={
  $1$, $4$, $16$, $64$, $256$
},
xmajorgrids,
y grid style={white!69.0196078431373!black},
ylabel={$t_{\textbf{wall}}/(N_{\textbf{steps}} N_{\textbf{dofs}})$  (s)},
ymin=8e-08, ymax=5e-05,
ymode=log,
ytick style={color=black},
ytick={1e-08,1e-07,1e-06,1e-05,0.0001,0.001},
yticklabels={
  \(\displaystyle {10^{-8}}\),
  \(\displaystyle {10^{-7}}\),
  \(\displaystyle {10^{-6}}\),
  \(\displaystyle {10^{-5}}\),
  \(\displaystyle {10^{-4}}\),
  \(\displaystyle {10^{-3}}\)
},
ymajorgrids,
]

\addplot [color0, mark=x, mark size=3, mark options={solid,rotate=270}, line width=1]
table{%
x  y
1 3.099794525299072e-06
2 1.6428876683044435e-06
4 8.812902671051025e-07
8 5.210408660125732e-07
16 2.757101873779297e-07
32 1.5306331199645995e-07
64 1.1201012702941896e-07
128 1.0924989524841309e-07
256 3.196120877838135e-07
};
\addlegendentry{\texttt{IMEX-o2}}

\addplot [color1,  mark=*, mark size=2.5, mark options={solid}, line width=1]
table{%
x  y
4   6.947471662521361e-06
8   3.6371219032287597e-06
16  1.9058574329376223e-06
32  1.200909992980957e-06
64  1.076472459411621e-06
128 9.526755210876466e-07
256 1.4095161880493165e-06
};
\addlegendentry{\dSDC (PinT)}

\addplot [color2, dashed, mark=square*, mark size=2, mark options={solid}]
table{%
x  y
1 2.0157207649230957e-05
2 1.0620718952941894e-05
4 5.823563481140137e-06
8 3.2958284534454348e-06
16 1.822273131942749e-06
32 1.0663395027160645e-06
64 9.235961059570312e-07
128 9.403191169738769e-07
256 1.8743259807586669e-06
};
\addlegendentry{\dSDC}
%


\end{axis}

\end{tikzpicture}%
\begin{tikzpicture}

\definecolor{color0}{rgb}{0.12156862745098,0.466666666666667,0.705882352941177}
\definecolor{color1}{rgb}{1,0.498039215686275,0.0549019607843137}
\definecolor{color2}{rgb}{0.172549019607843,0.627450980392157,0.172549019607843}
\definecolor{color3}{rgb}{0.83921568627451,0.152941176470588,0.156862745098039}
\definecolor{color4}{rgb}{0.580392156862745,0.403921568627451,0.741176470588235}

\begin{axis}[
legend cell align={left},
width=.49\textwidth,
legend style={
  fill opacity=0.8,
  draw opacity=1,
  text opacity=1,
  at={(-0.3,1.05)},
  anchor=south west,
  draw=white!80!black
},
log basis x={10},
log basis y={10},
tick align=outside,
tick pos=left,
x grid style={white!69.0196078431373!black},
xlabel={Number of \OpenMP threads},
xmin=0.7, xmax=320,
xmode=log,
xtick style={color=black},
xtick={1, 4, 16, 64, 256},
xticklabels={
  $1$, $4$, $16$, $64$, $256$
},
xmajorgrids,
y grid style={white!69.0196078431373!black},
ymin=8e-08, ymax=5e-05,
ymode=log,
ytick style={color=black},
ytick={1e-08,1e-07,1e-06,1e-05,0.0001,0.001},
yticklabels={
  \(\displaystyle {10^{-8}}\),
  \(\displaystyle {10^{-7}}\),
  \(\displaystyle {10^{-6}}\),
  \(\displaystyle {10^{-5}}\),
  \(\displaystyle {10^{-4}}\),
  \(\displaystyle {10^{-3}}\)
},
ymajorgrids,
]

\addplot [color0, mark=x, mark size=3, mark options={solid,rotate=270}, line width=1]
table{%
x  y
1 2.283779689025879e-06
2 1.1674545672607422e-06
4 6.531610583496094e-07
8 3.5696256317138676e-07
16 2.3373544342041015e-07
32 1.5391685546875e-07
64 1.7432111663818362e-07
128 3.227865216064453e-07
256 7.131554101562501e-07
};

\addplot [color1,  mark=*, mark size=2.5, mark options={solid}, line width=1]
table{%
x  y
4   5.103133282470703e-06
8   2.575070728302002e-06
16  1.586501037597656e-06
32  1.135307283782959e-06
64  1.0891937576293946e-06
128 9.356203201293945e-07
256 2.5777636695861815e-06
};

\addplot [color2, dashed, mark=square*, mark size=2, mark options={solid}]
table{%
x  y
1 1.3283182587432862e-05
2 7.799357251739502e-06
4 4.379822203826904e-06
8 2.4190981666564943e-06
16 2.1550899795532227e-06
32 1.7726692352294924e-06
64 1.2536278953552247e-06
128 1.8387457588195802e-06
256 4.742535327911377e-06
};
\end{axis}

\end{tikzpicture}%
	\caption{Strong scaling tests for \SWEET.\\
		Left: using a larger problem size $N_\text{dofs}=512^2=262144$ with \SWEET.\\
		Right: using a smaller problem size $N_\text{dofs}=256^2=65536$ with \SWEET.\\
		This figure shows \textit{only the performance of a single time step} with the possibility of SDC method to take larger time steps not taken into account.}
	\label{fig:strong_scaling_SWEET}
\end{figure}
We use a time step of $180$s for \dSDC,
and for \texttt{IMEX-o2} we use a time step of $75$s.
As for the result with \ICON, we show in \cref{fig:strong_scaling_SWEET} the ``time-per-DoF update'' against \OpenMP threads for \SWEET.

Again, scaled runtime is observed for two spatial grid sizes,
one with a reference resolution for the problem of interest (\cref{fig:strong_scaling_SWEET} left), and one with a coarser resolution for comparison (\cref{fig:strong_scaling_SWEET} right).

Here, space-parallelization only saturates around 64 cores
for the large problem, which appears to be a hardware limit as each algorithm
degrades its efficiency above that number of threads.
This is not the case for the smaller problem, as space-parallelization
saturates around 32 cores.
Hence space parallelization in \SWEET scales quite well as long as
the number of DoF per thread stays high enough,
due to the high efficiency of the
\texttt{SHTns} library \cite{schaeffer_efficient_2013} used in the code.
In that case, time parallelization for \dSDC has very little benefit over exclusive space parallelization,
even if the parallel speedup using only 4 time-parallel threads ($S\simeq 2.9$)
is still close to the theoretical speedup predicted by our model
($S_{theory} = 3.25$), albeit less close than it was for \ICON.
However, for the smaller problem size, time-parallelization of
\dSDC still manages to extend the scalability further up to
128 threads, where space-time parallelization allows a greater reduction
of wall-clock time compared to space-parallel only \dSDC.
This underlines the capability of time-parallelization to extend the parallel
efficiency of the solver only when space-parallelization saturates,
which is not completely the case for the large problem.
Finally, using more \OpenMP threads than physical cores available
(256 threads) degrades a lot the parallel performance in each configuration.

We also compare with the scaled wall-clock time of the reference time integration scheme, i.e., \texttt{IMEX-o2} for \SWEET.
The faster configuration for \dSDC is 8.5 times slower than \texttt{IMEX-o2}
for the larger problem, while only 6 times slower for the smaller problem,
illustrating again the better efficiency of space-time parallelization
for configuration where space-parallelization tends to saturates.
But as for \ICON, \dSDC using \texttt{MIN-SR-FLEX} coefficients
allows to take larger time-steps than \texttt{IMEX-o2}, while keeping
similar accuracy.
Whether it allows to reduce the computation time for a given time-window or not will be investigated in the next section.
\subsubsection{Work precision}
\Cref{fig:SWEET_work_precision_speedup} (left) shows work (measured in runtime) against precision (measured in absolute $L^\infty$-$L^2$ error of the vorticity) for $64$ \OpenMP threads for \SWEET.
As reference solution, we use the semi-implicit time integration scheme
of order 2 described previously, with a time-step of $1.5$s.
For \SWEET, \texttt{IMEX-o2} is more efficient for accuracies above $10^{-4}$.
Below that, first \dSDC and then time-parallel \dSDC become the most efficient
methods.
While parallel \dSDC provided shorter time-to-solution for \ICON for
medium to high accuracies, this is not the case for \SWEET anymore.
Indeed for the considered problem size ($N_\text{dofs}=512^2$)
the use of time-parallel \dSDC provides no improvement
to \dSDC (except for one point where time-parallel \dSDC is slightly more efficient),
as the problem size is big enough and space parallelization scales well.
But when reducing the problem size ($N_\text{dofs}=256^2$),
space-time parallelization allows using
more efficiently the parallel resources, as the scaling of space
parallelization degrades but not the one of time parallelization
(cf. Figure \ref{fig:strong_scaling_SWEET} (right),
associated work-precision plot is not given here).

\subsubsection{Speedup}
For \SWEET, speedups are shown in \cref{fig:SWEET_work_precision_speedup} (right) for methods calibrated for an absolute
accuracy of approximately $1.5\cdot10^{-5}$.
Again, using SDC provides some speedup over the IMEX base method,
due to the higher efficiency of \dSDC at this level of accuracy.
Here, however, there is little difference in performance between using space-only loop parallelization (\dSDC) or a nested space-time (\dSDC with PinT) approach.
Both cases provide a maximum speedup of around 32 for 64 threads with minimal difference in performance,
which is due to the space parallelization in \SWEET scaling better for this space grid size
(cf. discussion related to \cref{fig:strong_scaling_SWEET}).

For both frameworks \ICON and \SWEET
(cf. \cref{fig:ICON_work_precision_speedup,fig:SWEET_work_precision_speedup})
the optimal speedup is obtained when the number of threads is equal
to the number of physical cores,
which also indicates that hyper-threading does not improve the performance.

\begin{figure}[t]
	\centering
	\begin{tikzpicture}

\definecolor{color0}{rgb}{0.12156862745098,0.466666666666667,0.705882352941177}
\definecolor{color1}{rgb}{1,0.498039215686275,0.0549019607843137}
\definecolor{color2}{rgb}{0.172549019607843,0.627450980392157,0.172549019607843}

\begin{axis}[
	legend cell align={left},
	width=.48\textwidth,
	legend style={
	    fill opacity=0.8,
	    draw opacity=1,
	    text opacity=1,
	    at={(-0.3,1.05)},
	    anchor=south west,
	    draw=white!80!black
	},
	log basis x={10},
	log basis y={10},
	tick align=outside,
	tick pos=left,
	x grid style={white!69.0196078431373!black},
	xlabel={Wall-clock time (s)},
	xmajorgrids,
	xmin=176.123483694211, xmax=5400,
	xmode=log,
	xtick style={white!69.0196078431373!black},
	xtick={231,647,1626,4080},
	xticklabels={$231$,$647$,$1626$,$4080$},
	y grid style={gray},
	ylabel={absolute $L^\infty(0,T,L^2(\Omega))$},
	ymajorgrids,
	ymin=1e-7, ymax=0.000553010675504651,
	ymode=log,
	ytick style={color=black},
	ytick={1e-09,1e-08,1e-07,1e-06,1e-05,0.0001,0.001,0.01},
	yticklabels={
	    \(\displaystyle {10^{-9}}\),
	    \(\displaystyle {10^{-8}}\),
	    \(\displaystyle {10^{-7}}\),
	    \(\displaystyle {10^{-6}}\),
	    \(\displaystyle {10^{-5}}\),
	    \(\displaystyle {10^{-4}}\),
	    \(\displaystyle {10^{-3}}\),
	    \(\displaystyle {10^{-2}}\)
}
]

\addplot [color0, mark=x, mark size=3, mark options={solid,rotate=270}, line width=1]
table {%
231.62397388	0.0002221206569140285
316.909525062	0.0001243551712087769
459.839344554	5.4991832249937615e-05
993.521747378	1.370187808816003e-05
4081.58644619	8.552229093520179e-07
};

\addplot [color1,  mark=*, mark size=2.5, mark options={solid}, line width=1]		
table {%
309.239477142 0.00029285003923959415
343.460210928 0.0001761336471928683
748.675682761 1.6535340135096327e-05
1004.6606896 3.3293778714512108e-06
2303.43286138 2.01575578734151e-07
};

\addplot [color2, dashed, mark=square*, mark size=2, mark options={solid}]
table {%
256.034779955	0.00029285003923959415
290.170194622	0.0001761336471928683
647.3982542	1.6535340135096327e-05
1038.07210801	3.3293778714512108e-06
1626.49432032	2.01575578734151e-07
};

\end{axis}
    
\end{tikzpicture}%
	\begin{tikzpicture}
	
	\definecolor{color0}{rgb}{0.12156862745098,0.466666666666667,0.705882352941177}
	\definecolor{color1}{rgb}{1,0.498039215686275,0.0549019607843137}
	\definecolor{color2}{rgb}{0.172549019607843,0.627450980392157,0.172549019607843}
	
	\begin{axis}[
		legend cell align={left},
		width=.49\textwidth,
		legend style={
			fill opacity=0.8,
			draw opacity=1,
			text opacity=1,
			at={(0.99,0.01)},
			anchor=south east,
			draw=white!80!black
		},
		log basis x={10},
		log basis y={10},
		tick align=outside,
		tick pos=left,
		x grid style={white!69.0196078431373!black},
		xlabel={Number of \OpenMP threads},
		xmin=0.795951193613564, xmax=300,
		xmode=log,
		xtick style={color=black},
		xtick={1,4,16,64,256},
		xticklabels={$1$,$4$,$16$,$64$,$256$},	
		xmajorgrids,
		y grid style={white!69.0196078431373!black},
		ylabel={Speedup},
		ymin=0.9, ymax=50,
		ymode=log,
		ytick style={color=black},
		ytick={1,2,4,8,16,32},
		ymajorgrids,
		yticklabels={1,2,4,8,16,32}
		]
		
		\addplot [color0, mark=x, mark size=3, mark options={solid,rotate=270}, line width=1]
		table{%
			x  y
			1.0 1.0
			2.0 1.8468907340242247
			4.0 3.3440666313248295
			8.0 6.008475280913207
			16.0 11.123683283408049
			32.0 19.800121935717762
			64.0 25.520015244123037
			128.0 26.760825352405163
			256.0 10.193793439612419
		};
		\addlegendentry{{\texttt{IMEX-o2}}}
		\addplot [color1,  mark=*, mark size=2.5, mark options={solid}, line width=1]		
		table{%
			x  y
			4.0 5.38098586816197
			8.0 10.211790873071992
			16.0 19.023408067826626
			32.0 29.893755481811567
			64.0 35.85216114514842
			128.0 36.447844988435314
			256.0 23.708297479448415
		};
		\addlegendentry{\dSDC (PinT)}
		\addplot [color2, mark=square*, dashed, mark size=2, mark options={solid}, line width=1]
		table{%
			x  y
			1.0 1.7790256471551487
			2.0 3.406654597481127
			4.0 6.040425950274304
			8.0 10.735996002878052
			16.0 20.399024564097207
			32.0 30.924112842449965
			64.0 37.60517798911689
			128.0 37.4274876416278
			256.0 21.838545647695945
		};
		\addlegendentry{\dSDC}
	\end{axis}
	
\end{tikzpicture}%
	\caption{Left: Work-precision plot \SWEET, using $64$ threads for IMEX and \dSDC, and nested parallelization with $32$ threads in space and $4$ threads in time for \dSDC (PinT). Various time-step sizes are used, for IMEX $\dt \in \{60,\dots, 3.75\}$ (s)	and for \dSDC $\dt \in \{450,\dots, 60\}$ (s).
		We observe hardly any performance gains for SWEET.
		This can be accounted for by the very good scaling behavior of SWEET.
		A smaller problem size would lead again to speedups for SWEET (not shown here).\\
		Right: Speedup of \dSDC for \SWEET, comparing IMEX ($\dt=15$s) and \dSDC ($\dt=180$s) at similar absolute accuracy of $\approx 1.5\cdot10^{-5}$. }
	\label{fig:SWEET_work_precision_speedup}
\end{figure}


\section{Discussion and outlook}
This paper presents a shared-memory parallel implementation of parallel-across-the-methods spectral deferred corrections.
For the first time, we demonstrate parallel speedups for two established simulation codes.
In the operational ocean model \ICON, a modified SDC implementation is used, adapted to the existing time-stepping infrastructure, whereas the standard formulation of SDC is implemented in the research code \SWEET.
For both implementations, theoretical considerations predict a speedup of around $3.25$ over serial SDC.
In our implementations we reach speedups of $3.4$ for \ICON and $2.9$ for \SWEET.
In both codes, SDC delivers shorter time-to-solution than the standard AB/IMEX methods for medium to high accuracies simply due to the higher-order.
Furthermore, we demonstrate that a nested \OpenMP implementation in \ICON with parallel SDC as outer and multi-threading of mesh operations as inner loop increases speedup significantly.
By contrast, in \SWEET, space-only and space-time parallelization perform almost identically.
This provides insight that benefits gained from parallel SDC do heavily depend not only on implementation but also on the used spatial discretization.
While parallel SDC requires more investigation to accelerate (pseudo)-spectral methods like the ones used in \SWEET, it seems to be already suited to work in combination with mesh-based finite volume methods used in \ICON.
Hence, a time-parallel \dSDC implementation seems to be useful for large code already space-parallelized to squeeze out additional speedup without performing a major refactoring.

For \ICON, the next step is to combine the proposed parallel-in-time scheme with the existing \MPI parallelization.
The parallel efficiency of our \dSDC integrator coupled with a distributed spatial solver is likely to improve over a sequential time integrator with a strictly distributed spatial solver since the \MPI communication overhead will be reduced.
Also, this advantage should be maintained as long
as computation time for tendency evaluation and CG time dominates,
which is usually the case for large scale simulations.
Further hybrid parallelism will allow scalability to climatologically relevant scales.
Also, the recent trend of transferring computations to GPUs is interesting with regard to our parallelization approach especially since most of the \ICON code has already been ported to GPUs.
Therefore, a fully GPU-based space-time parallelization may be of interest for future research.
For \SWEET, further variants of \dSDC and beyond are planned to be systematically explored:
in particular,
the use of other diagonal coefficients for \dSDC to improve the numerical stability, or changing the tendency terms integrated implicitly or explicitly
to investigate other configuration that could lead to more efficient approaches with \dSDC.

Moreover, the theoretical understanding of \dSDC may improve further since we mostly adopted parameters
that were derived for purely implicit schemes.
Hence, further improvements to the optimized parameters may be possible for the IMEX-type approach used here.


\begin{credits}
\subsubsection{\ackname} 
The authors gratefully acknowledge the Gauss Centre for Supercomputing 
e.V. (www.gauss-centre.eu) for funding this project by providing computing time 
through the John von Neumann Institute for Computing (NIC) on the GCS Supercomputer JUWELS \cite{JUWELS},
and the SimLab Climate project for providing computing time on JUSUF \cite{JUSUF},
both clusters hosted at Jülich Supercomputing Centre (JSC).

\subsubsection{\discintname}
The authors have no competing interests to declare that are relevant to the content of this article.
\end{credits}
\bibliographystyle{splncs04}
\bibliography{mybibliography,pint}

\end{document}